\documentclass[preprint2]{aastex6}
\usepackage{cancel}	
\usepackage{color}
\usepackage{apjfonts}
\usepackage{affils}
\usepackage{graphicx,subfigure}
\usepackage{upgreek}
\usepackage{microtype}

\newcommand{\kms}{km\,s$^{-1}$}
\def\cmg{cm$^2$\,g$^{-1}$}

\def\M{M$_{\odot}$}

\def\Hb{H{$\beta$}}

 \def\Mej{$M_{\rm ej}$}

\def\tsd{$\tau_{\rm sd}$}

\begin{document}


\title{An ultraviolet excess in the superluminous supernova Gaia16apd reveals a powerful central engine}

\shortauthors{Nicholl et al.}

\shorttitle{Extreme UV emission from Gaia16apd}


\DeclareAffil{cfa}{Harvard-Smithsonian Center for Astrophysics, 60 Garden Street, Cambridge,
  Massachusetts 02138, USA; \href{mailto:matt.nicholl@cfa.harvard.edu}{matt.nicholl@cfa.harvard.edu}}
\DeclareAffil{queens}{Astrophysics Research Centre, School of Mathematics and Physics, Queens University Belfast, Belfast BT7 1NN, UK}
\DeclareAffil{nyu}{Center for Cosmology and Particle Physics, New York University, 4 Washington Place, New York, NY 10003, USA}
\DeclareAffil{mpe}{Max-Planck-Institut f{\"u}r Extraterrestrische Physik, Giessenbachstra\ss e 1, 85748, Garching, Germany}
\DeclareAffil{mpa}{Max-Planck-Institut f{\"u}r Astrophysik, Karl-Schwarzschild-Str. 1, D-85748 Garching, Germany}
\DeclareAffil{lcogt}{Las Cumbres Observatory Global Telescope, 6740 Cortona Dr, Suite 102, Goleta, CA 93111, USA}
\DeclareAffil{kitp}{Kavli Institute for Theoretical Physics, University of California, Santa Barbara, CA 93106, USA}
\DeclareAffil{soton}{School of Physics and Astronomy, University of Southampton, Southampton, SO17 1BJ, UK}
\DeclareAffil{hawaii}{Institute for Astronomy, University of Hawaii at Manoa, Honolulu, HI 96822, USA}
\DeclareAffil{ohio}{Astrophysical Institute, Department of Physics and Astronomy, 251B Clippinger Lab, Ohio University, Athens, OH 45701, USA}
\DeclareAffil{cambridge}{Institute of Astronomy, University of Cambridge, Madingley Road, Cambridge, CB3 0HA}
\DeclareAffil{weizmann}{Benoziyo Center for Astrophysics, Weizmann Institute of Science, Rehovot 76100, Israel}
\DeclareAffil{millennium}{Millennium Institute of Astrophysics, Vicu\~{n}a Mackenna 4860, 7820436 Macul, Santiago, Chile}
\DeclareAffil{chile}{Departamento de Astronom\'ia, Universidad de Chile, Camino El Observatorio 1515, Las Condes, Santiago, Chile}
\DeclareAffil{tuorla}{Tuorla Observatory, Department of Physics and Astronomy, University of Turku, V\"ais\"al\"antie 20, FI-21500 Piikkiö, Finland}
\DeclareAffil{osu}{Department of Astronomy, The Ohio State University, 140 West 18th Avenue, Columbus, OH 43210, USA}
\DeclareAffil{ccapp}{Center for Cosmology and AstroParticle Physics (CCAPP), The Ohio State University, 191 W. Woodruff Ave., Columbus, OH 43210, USA}
\DeclareAffil{ucsb}{Department of Physics, University of California, Santa Barbara, Broida Hall, Mail Code 9530, Santa Barbara, CA 93106-9530, USA}
\DeclareAffil{taiwan}{Institute of Astronomy, National Central University, Chung-Li 32054, Taiwan}
\DeclareAffil{columbia}{Columbia Astrophysics Laboratory, Columbia University, New York, NY 10027, USA}
\DeclareAffil{sorbonne}{Sorbonne Universit\'es, UPMC, Paris VI,UMR 7585, LPNHE, F-75005, Paris, France}
\DeclareAffil{cnrs}{CNRS, UMR 7585, Laboratoire de Physique Nucleaire et des Hautes Energies, 4 place Jussieu, 75005 Paris, France}
\DeclareAffil{not}{Nordic Optical Telescope, Apartado 474, E-38700 Santa Cruz de La Palma, Spain}
\DeclareAffil{carnegie}{Carnegie Observatories, 813 Santa Barbara Street, Pasadena, CA 91101, USA}
\DeclareAffil{hcpf}{Hubble, Carnegie-Princeton Fellow}
\DeclareAffil{davis}{Department of Physics, University of California, Davis, CA 95616, USA}
\DeclareAffil{mit}{Kavli Institute for Astrophysics and Space Research, Massachusetts Institute of Technology, 77 Massachusetts Avenue, Cambridge, MA 02139}
\DeclareAffil{caltech}{Astronomy Department, California Institute of Technology, Pasadena, California 91125, USA}
\DeclareAffil{jmu}{Astrophysics Research Institute, Liverpool John Moores University, IC2, Liverpool Science Park, 146 Brownlow Hill, Liverpool L3 5RF, UK}
\DeclareAffil{eso}{European Southern Observatory, Karl-Schwarzschild-Str. 2, D-85748 Garching, Germany}
\DeclareAffil{northwestern}{Center for Interdisciplinary Exploration and Research in Astrophysics (CIERA) and Department of Physics and Astronomy, Northwestern University, Evanston, IL 60208}

\affilauthorlist{M.~Nicholl\affils{cfa},
E.~Berger\affils{cfa},
R.~Margutti\affils{northwestern},
P.~K.~Blanchard\affils{cfa},
D.~Milisavljevic\affils{cfa},
P.~Challis\affils{cfa},
B.~D.~Metzger\affils{columbia},
R.~Chornock\affils{ohio}
}

\begin{abstract}

Since the discovery of superluminous supernovae (SLSNe) in the last decade, it has been known that these events exhibit bluer spectral energy distributions than other supernova subtypes, with significant output in the ultraviolet. However, the event Gaia16apd seems to outshine even the other SLSNe at rest-frame wavelengths below $\sim 3000$\,\AA. \citet{yan2016} have recently presented \textit{HST} UV spectra and attributed the UV flux to low {iron-group abundance in the outer ejecta}, and hence reduced line blanketing. Here we present UV and optical light curves over a longer baseline in time, revealing a rapid decline at UV wavelengths despite a typical optical evolution. Combining the published UV spectra with our own optical data, we demonstrate that Gaia16apd has a much hotter continuum than virtually any SLSN at maximum light, but it cools rapidly thereafter and is indistinguishable from the others by $\sim 10$--15 days after peak. Comparing the equivalent widths of UV absorption lines with those of other events, we show that the excess UV continuum is a result of a more powerful central power source, rather than a lack of UV absorption relative to other SLSNe or an additional component from interaction with the surrounding medium. These findings strongly support the central-engine hypothesis for hydrogen-poor SLSNe. An explosion ejecting $M_{\rm ej} = 4.8 (0.2/\kappa)$\,\M, where $\kappa$ is the opacity in \cmg, and forming a magnetar with spin period $P=2$\,ms, and $B=2\times10^{14}$\,G (lower than other SLSNe with comparable rise-times) can consistently explain the light curve evolution and high temperature at peak. The host metallicity, $Z=0.18$\,Z$_\odot$, is comparable to other SLSNe.

\end{abstract}

\keywords{supernovae: general --- supernovae: Gaia16apd}

\section{Introduction}\label{sec:intro}

Superluminous supernovae (SLSNe) reach luminosities of $\sim 10^{44}$\,erg\,s$^{-1}$, which is 10--100 times brighter than any previously known supernova (SN), and they are especially luminous in the UV \citep{qui2011}.
 While these explosions do seem to come from massive stars \citep{gal2012,nic2015b}, often following the loss of their hydrogen envelope \citep{ins2013}, the luminosity cannot be explained in the same way as for other core-collapse SNe: a shock depositing energy in the stellar envelope fails because the required efficiency is too large, and the mass of $^{56}$Ni needed to power the light curve through radioactive decay often exceeds the total mass budget of the explosion. Instead, the debate has centred around whether the excess energy is input from an \textit{external} source, such as the interaction of the SN ejecta with a massive shell of circumstellar material \citep[CSM, e.g.][]{che2011}, or an \textit{internal} engine, for which a highly magnetised neutron star remnant with a millisecond spin period seems to be the best candidate \citep{kas2010}.

Gaia16apd was discovered by the Photometric Science Alerts system from the \textit{Gaia} survey \citep{wyr2012}, and classified by the NOT Unbiased Transient Survey as a young, hydrogen-poor SLSN in a very faint galaxy at redshift $z=0.102$ \citep{kan2016}. This makes it the second-nearest SLSN discovered to date. It quickly became clear that proximity was not the only thing that was special about this event---it was also extraordinarily UV-bright, being $\sim1.5$ magnitudes brighter in the UV ($\sim2000-3000$\,\AA) than the next-nearest SLSN, PTF12dam \citep{nic2013}, despite a similar brightness in the optical. Recently, \citet{yan2016} presented a detailed analysis of the UV spectrum of Gaia16apd at maximum light, showing that it was subject to significantly less line-blanketing than any normal-luminosity SNe, and hence likely metal-poor in the outer ejecta. While this may indeed explain why SLSNe in general are more UV-luminous than normal SNe, it does not tell the full story of Gaia16apd. In this paper, we show how its UV and optical evolution fit into the context of other SLSNe, with particular emphasis on what the UV diversity of these events can teach us of the power source. We argue that the observed properties can be accounted for self-consistently by the popular magnetar model of SLSNe, and the UV excess in particular is a natural consequence of a short spin period and relatively low magnetic field in combination with a modest ejecta mass.

\section{Observations}\label{sec:obs}

\begin{figure}
\centering
\includegraphics[width=8.5cm]{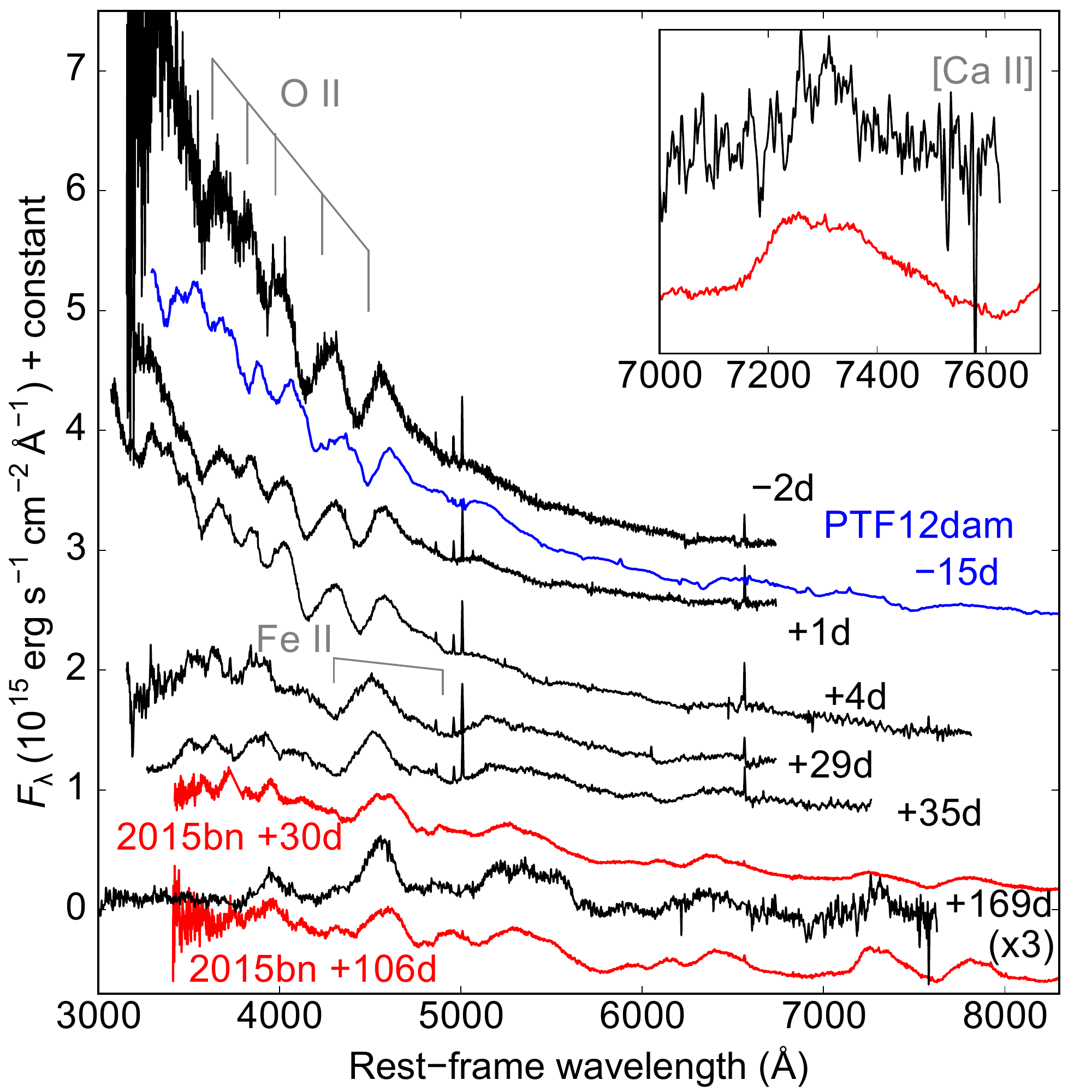}
\caption{\small Spectroscopy of Gaia16apd. The optical spectra of Gaia16apd obtained from FAST and MMT show a typical evolution from a blue continuum with broad lines of singly-ionised oxygen to a redder spectrum dominated by iron and intermediate mass elements. Two of the best-observed low-redshift SLSNe are shown for comparison. Labels indicate the time in days with respect to $r$-band maximum light, in the SN rest-frame. {The late-time spectrum shows [\ion{Ca}{2}] emission, previously seen in a number of slow-evolving SLSNe \citep{nic2016b,kan2016b}}.}
\label{fig:spec}
\end{figure}

When Gaia16apd was announced, we immediately triggered follow-up observations from ground-based observatories and the \textit{Swift} satellite. We obtained spectroscopic observations using the FAST and Blue Channel spectrographs on the 60'' and MMT telescopes, respectively, at Fred Lawrence Whipple Observatory (FLWO); FAST spectra were reduced using a dedicated pipeline, while Blue Channel data were reduced in \textsc{pyraf}. We show our spectra of Gaia16apd in Figure \ref{fig:spec}. We take the date of maximum light to be MJD 57541 for consistency with \citet{yan2016}.

The spectra are typical of hydrogen-poor SLSNe, as demonstrated by comparison to some of the best-observed events, showing the usual transition from a very blue spectrum with O\,\textsc{ii} lines to a redder spectrum resembling a normal-luminosity Type Ic SN \citep{pas2010}. From the narrow host galaxy lines visible in the spectra, we measure a redshift of $z=0.1013$, in good agreement with the original classification. We use this value for the redshift throughout. We correct only for Milky Way extinction \citep{schlaf2011}, as the host galaxy of Gaia16apd shows no evidence for a Balmer ratio in excess of case B recombination \citep{ost1989} nor significant \ion{Na}{1} absorption \citep{poz2012}.

Imaging observations were obtained in optical passbands from the 48'' imaging telescope at FLWO and de-biased/flat-fielded using \textsc{astropy} packages. Photometry was determined by point-spread function fitting with a zero point derived from local field stars; the magnitudes of these stars were taken from the Pan-STARRS 3$\pi$ survey where possible, or calibrated against standard fields on photometric nights in the case of $B$ and $V$ filters. Further photometry in the UV and optical were obtained with the UVOT instrument on \textit{Swift} (Cycle 12 GI program \#1215102) and extracted following \citet{bro2009}. Colour corrections\footnote{http://heasarc.gsfc.nasa.gov/docs/heasarc/caldb/swift/docs/\\uvot/uvot\_caldb\_coltrans\_02b.pdf} were applied to convert $u,b,v$ magnitudes to the more standard $U,B,V$ system. The earliest data points are from \textit{Gaia} and PTF. We converted the \textit{Gaia} $G$-band photometry to $i$-band using the observed $r-i$ colour and the post-launch colour-conversions from their Data Release 1\footnote{https://gaia.esac.esa.int/documentation/GDR1/\\Data\_processing/chap\_cu5phot/sec\_phot\_calibr.html\#SS5}. The PTF $g$-band point was presented by \citet{yan2016}. All data will be made available through the Open Supernova Catalog \citep{gui2016}.

\begin{figure*}
\centering
\includegraphics[width=8.cm]{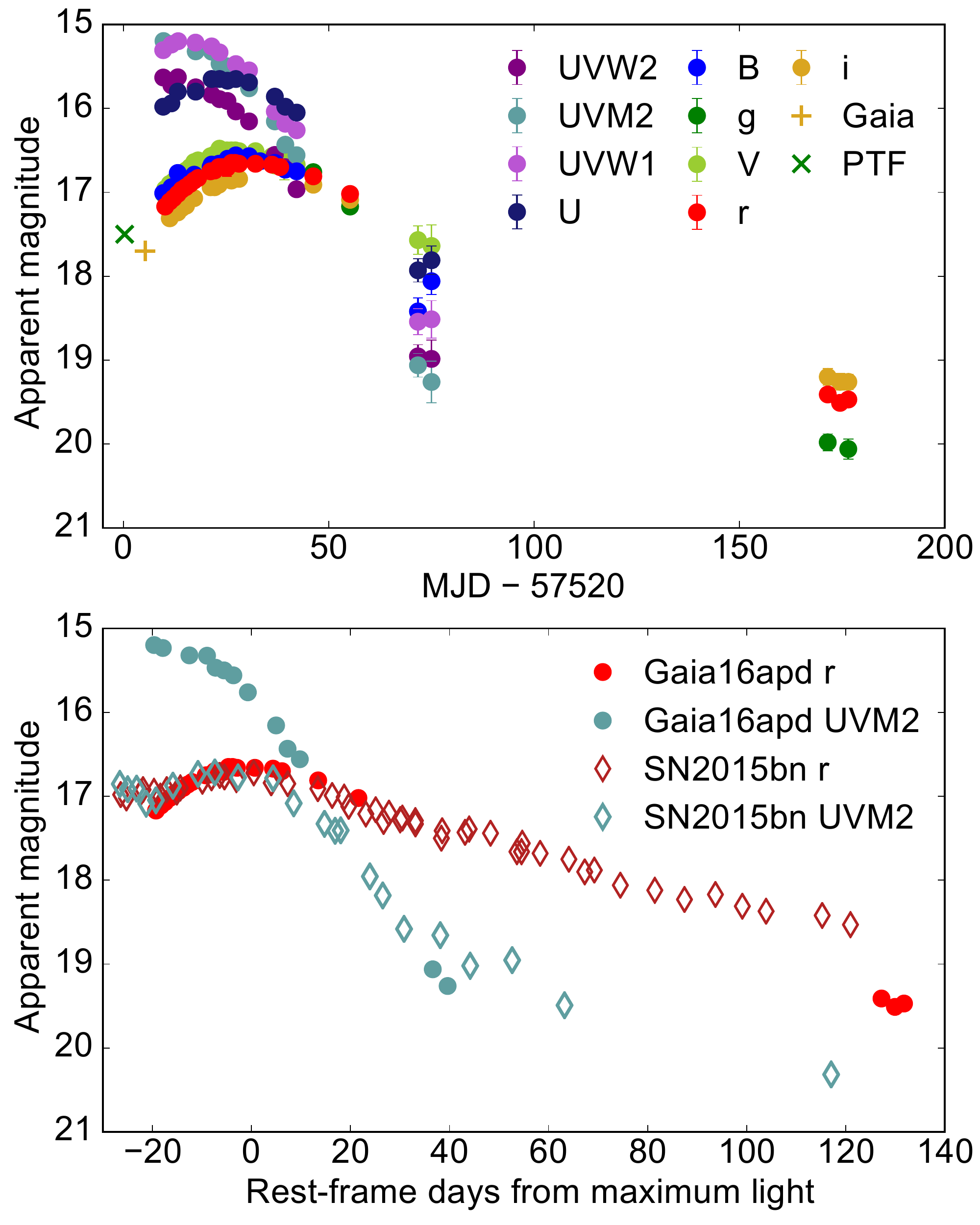}
\includegraphics[width=8.cm]{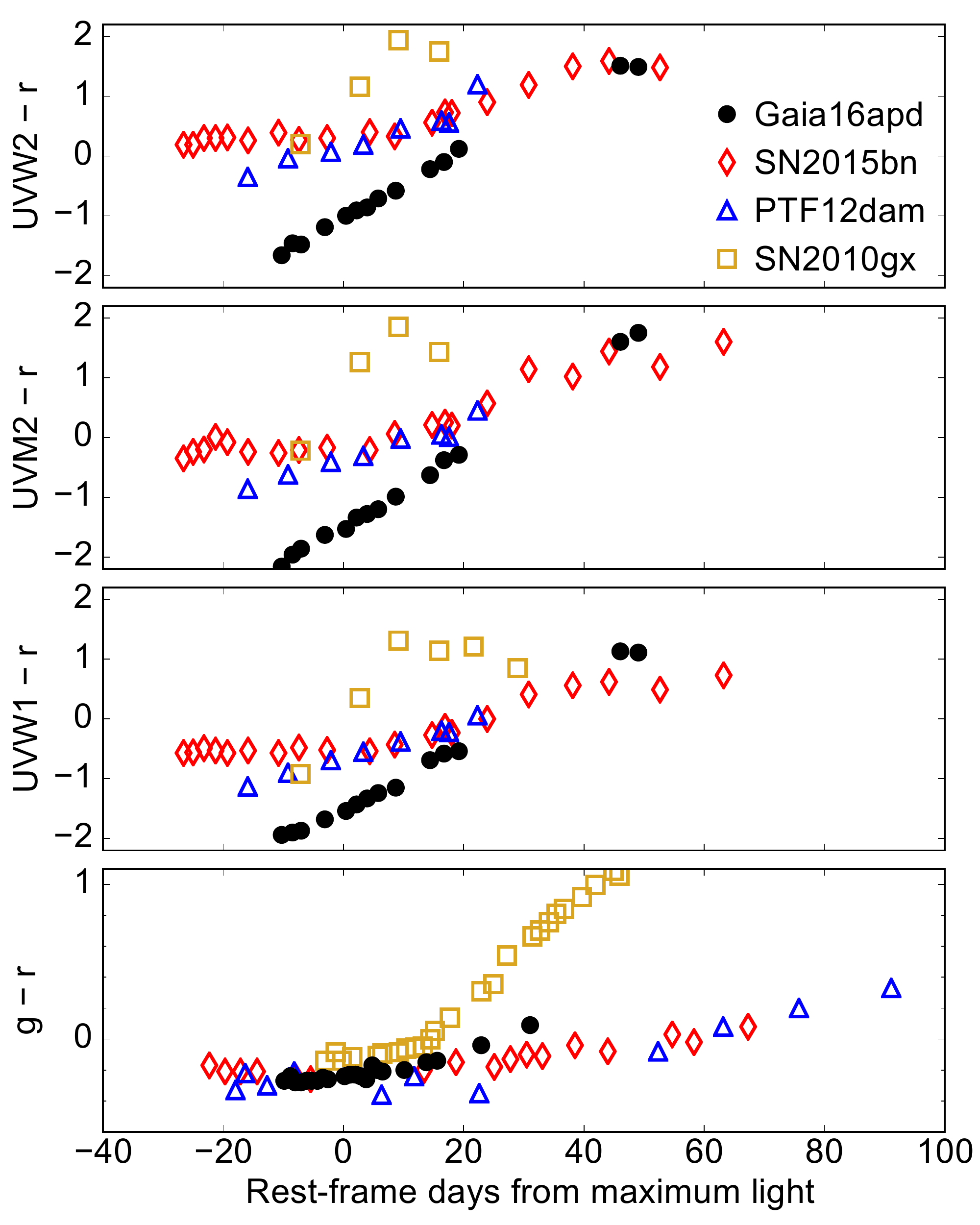}
\caption{\small Optical and UV photometry of Gaia16apd. Top left: Multicolour lightcurves. Bandpasses are labelled in order from bluest to reddest. Note the initial brightness and rapid decline at UV wavelengths. Lower left: Comparison with SN\,2015bn \citep{nic2015b}, the best-observed SLSN to date, which is at a similar redshift ($z=0.1136$) to Gaia16apd. Despite a similar $r$-band (optical) brightness, Gaia16apd is initially almost 2 magnitudes brighter in the UV. Right: Colour evolution of Gaia16apd compared to other low-$z$ SLSNe. The \textit{UV}\,$-$\,\textit{optical} colours redden from their extreme early values to look normal by 1--2 weeks after optical maximum light. Gaia16apd displays a typical $g-r$ colour evolution, as may be expected on the Rayleigh-Jeans tail of the SED.
}
\label{fig:phot}
\end{figure*}

Our multicolour light curves are shown in Figure \ref{fig:phot}, highlighting the remarkable UV excess. We compare the colour evolution to other events with well-sampled UV light curves from \textit{Swift}. PTF12dam \citep{nic2013} and SN\,2015bn \citep{nic2016b} are both at similar redshift to Gaia16apd ($z\approx0.1$), however we caution that SN\,2010gx, at $z=0.23$ \citep{pas2010}, may be subject to a significant $K$-correction that is difficult to evaluate without UV spectra. The \textit{UV}\,$-$\,\textit{optical} colours of Gaia16apd are bluer by 1--2 magnitudes at 20 days before $r$-band maximum, but evolve quickly such that by 10--15 days after maximum, all of the $z\approx0.1$ events show near-identical colours. The convergence in colour suggests that the extinction in the host galaxy is indeed low, or at least similar to that in other SLSNe.

The optical $g-r$ colour evolution is perfectly consistent with the other events, demonstrating that if we had observed Gaia16apd only in the optical (or began UV follow-up shortly after maximum light), it would have looked like an entirely typical SLSN. The only other claimed SLSN with such an extreme \textit{UV}\,$-$\,\textit{optical} colour is ASASSN-15lh \citep{dong2016}. However, the interpretation of that event is contentious, and the spectroscopic evolution does not resemble the classic SLSN sequence shown in Figure \ref{fig:phot} \citep{lel2016,mar2016}. Thus Gaia16apd represents the first spectroscopically-normal SLSN to display this copious but rapidly-fading UV emission. 

\begin{figure*}
\centering
\includegraphics[width=8.cm]{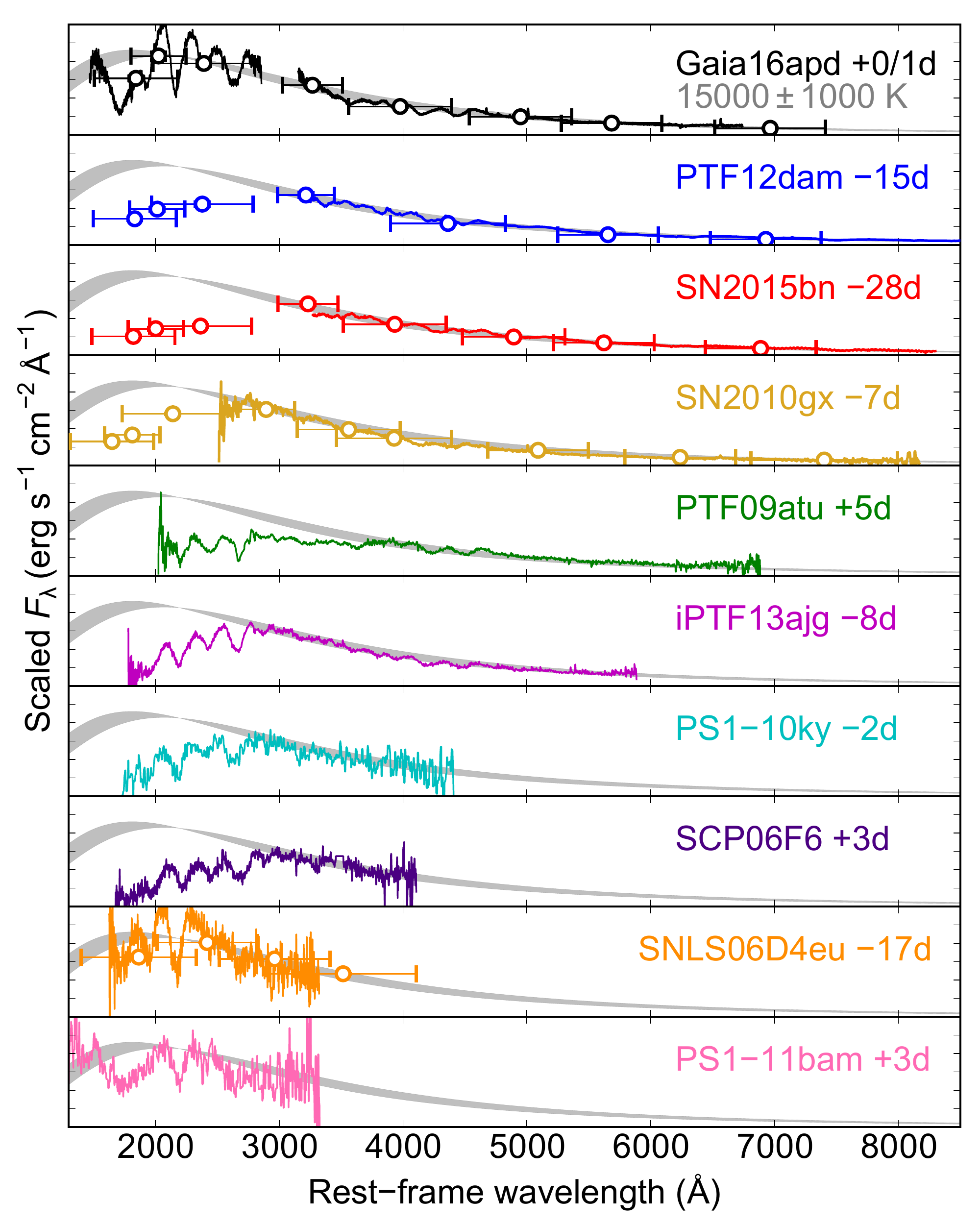}
\includegraphics[width=8.cm]{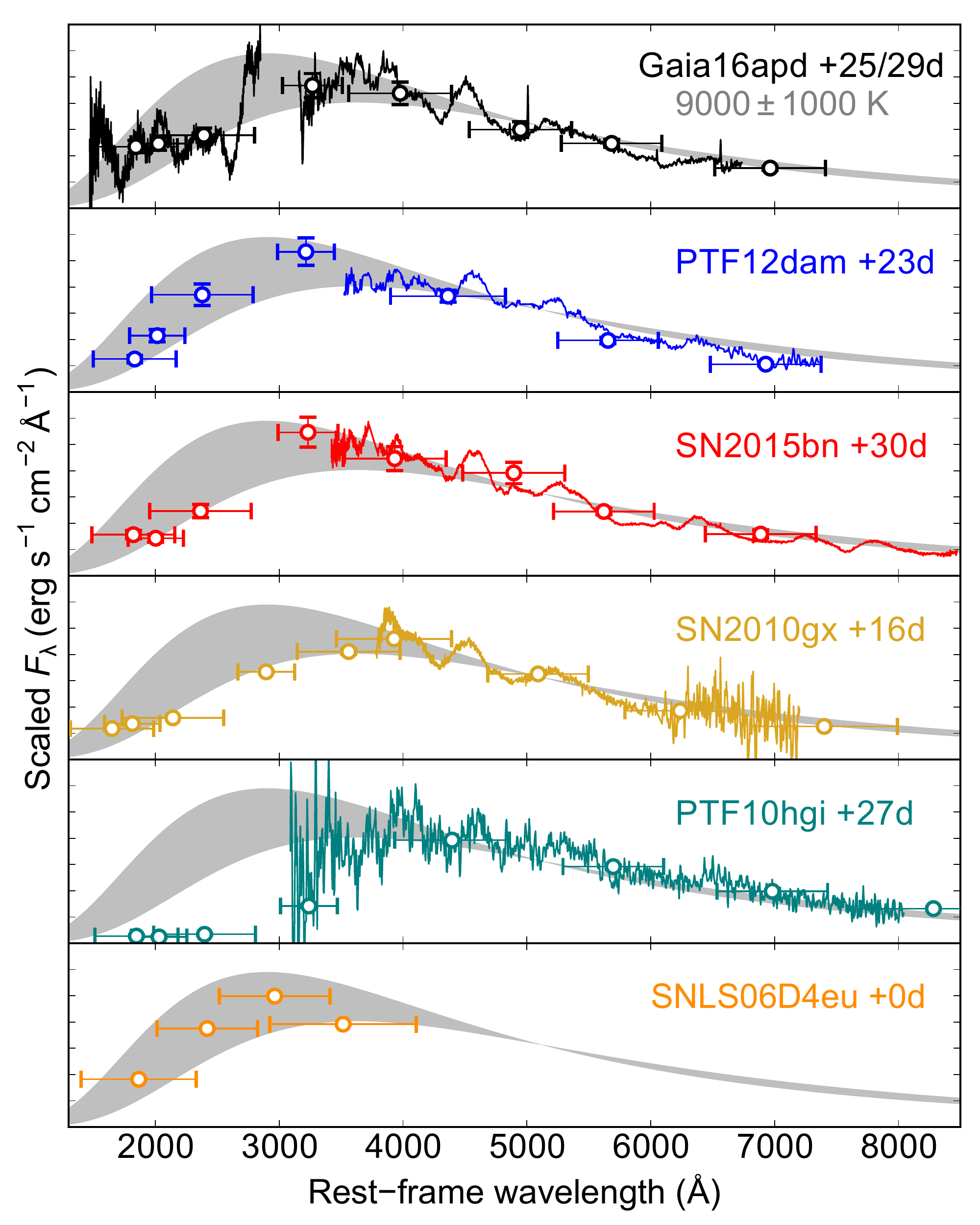}
\caption{\small Left: The UV-optical SEDs of all SLSNe with broad wavelength coverage before or around maximum light. A 15000\,K blackbody SED is shown for comparison. While this gives a good match to all SLSNe at wavelengths $\ga 3000$\,\AA, most objects show a significant deficit in the UV compared to the blackbody. The photometry of Gaia16apd and PS1011bam at maximum light, and SNLS06d4eu at very early times, are consistent with the blackbody curve, while their spectra suggest an even higher temperature. Right: The same comparison at several weeks after maximum (or at maximum for SNLS06d4eu). This time a blackbody of 8000--10000\,K gives a reasonable approximation to the flux levels across the whole UV-optical regime for most SLSNe, including Gaia16apd. Data are from \citet{nic2013,nic2016b,pas2010,qui2011,vre2014,chom2011,bar2009,how2013,ins2013}.}
\label{fig:sed}
\end{figure*}

To demonstrate this more concretely, we plot a comparison of the spectral energy distribution (SED) for all SLSNe with available UV data in Figure \ref{fig:sed}, separating the data into early- and late-time observations.
Gaia16apd was observed spectroscopically in the UV with the \textit{Hubble Space Telescope} (\textit{HST}) by \citet{yan2016}. The earliest UV spectrum was obtained on MJD 57541, approximately the time of bolometric maximum light, and thus during the UV-luminous phase, and further spectra were obtained at 11 and 25 days later. We downloaded these spectra from the \textit{HST} archive and overplot them with the SED derived from our \textit{Swift} photometry. The flux calibration is consistent between the spectra and photometry.

It is clear that most SLSNe display significant flux suppression below $\sim 3000$\,\AA\ (though as pointed out by others, e.g.~\citealt{qui2011,yan2016}, less so than normal-luminosity SNe). This is true despite many of these objects being observed much earlier than Gaia16apd with respect to maximum light; early observations would favour bluer SEDs, as SNe almost universally cool with time. {Two SLSNe that do show notably blue SEDs are SNLS06d4eu \citep{how2013} and PS1-11bam \citep{ber2012}. However, at $z\approx1.6$, almost all of the data for these events are rest-frame UV rather than optical, and we lack times-series spectra. While SNLS06d4eu was only observed spectroscopically at $-17$\,d from peak, PS1-11bam shows a blue UV spectrum at maximum light, similar to Gaia16apd as pointed out by \citet{yan2016}. If this event had optical data, we likely would have seen a very blue  \textit{UV}\,$-$\,\textit{optical} colour like that in Gaia16apd.}

For the late time comparison, we use the combined UV-optical SED at $+25$--29\,d, scaled to the photometry at +25\,d. The similarity to PTF12dam and SN\,2015bn at this phase agrees with our colour comparison, and shows that the UV properties of Gaia16apd are no longer out of the ordinary. The UV SED of SNLS06d4eu at maximum light matches other objects at later phases, showing that this event does not remain exceptionally UV-luminous for as long as Gaia16apd. PTF10hgi \citep{ins2013} seems to be particularly UV-faint for a SLSN.

\section{The nature of the UV excess}\label{sec:dis}

We have so far demonstrated that Gaia16apd shows a pronounced UV excess at maximum light relative to other SLSNe, but that this emission fades to a more typical level shortly afterwards. Determining the source of the UV emission from Gaia16apd thus provides a new means to probe the physics of SLSNe. 

\citet{yan2016} propose that low {iron-group abundance} is a key factor in explaining the copious UV flux in Gaia16apd. {This could result from either a low natal metallicity or a lack of heavy-element synthesis in the explosion}.
Using our three latest (deepest) spectra, we estimate the metallicity of the host galaxy from measured line ratios (\Hb, [\ion{O}{2}], [\ion{O}{3}]). The common $R_{23}$ diagnostic \citep[assuming the lower branch;][]{kob1999}, calibrated in the \citet{mcg1991} scale, gives $12+\log({\rm O}/{\rm H})=7.94\pm0.06$, or $Z=0.18$\,Z$_\odot$. While this is at the low end for SLSN hosts, it is similar to various objects from \citet{lun2014,chen2015,lel2015,per2016}, and specifically the hosts of SN\,2010gx \citep{chen2013} and SN\,2015bn \citep{nic2016b}. \citet{yan2016} also note that the luminosity (a proxy for metallicity) of this galaxy in archival imaging is similar to other SLSN hosts.
{More generally, none of the synthetic SLSN spectra in Figure 8 of \citet{maz2016}, which span an order of magnitude in metal abundance above the photosphere, come close to reproducing the observed UV flux levels of Gaia16apd}. Thus it seems very unlikely that metal abundance alone can account for the UV excess {relative to other SLSNe}.

One possibility is a short-lived additional energy source. In this interpretation, a high temperature would be required for the extra component such that most of the energy is emitted in the UV, which could point towards shock heating. While the timescale of the observed UV excess (a few weeks) is much too long for post-shock cooling of the stellar envelope---which expands rapidly and degrades the thermal energy content adiabatically---a shock passing through an extended CSM could generate a longer-lived high-temperature component. The other possibility is that there is only a single component to the luminosity, which peaks in the UV at early times and still manages to produce an optically-normal SLSN. We compare our multicolour photometry to blackbody models and do not find evidence for separate components with different temperatures (Figure \ref{fig:sed}). We therefore favour a single power source for the UV and optical emission. This requires that Gaia16apd has a hotter continuum temperature than other SLSNe at a similar phase from maximum light. Determining what sets this temperature will have important implications for understanding the underlying power source.

\begin{figure}
\centering
\includegraphics[width=8.cm]{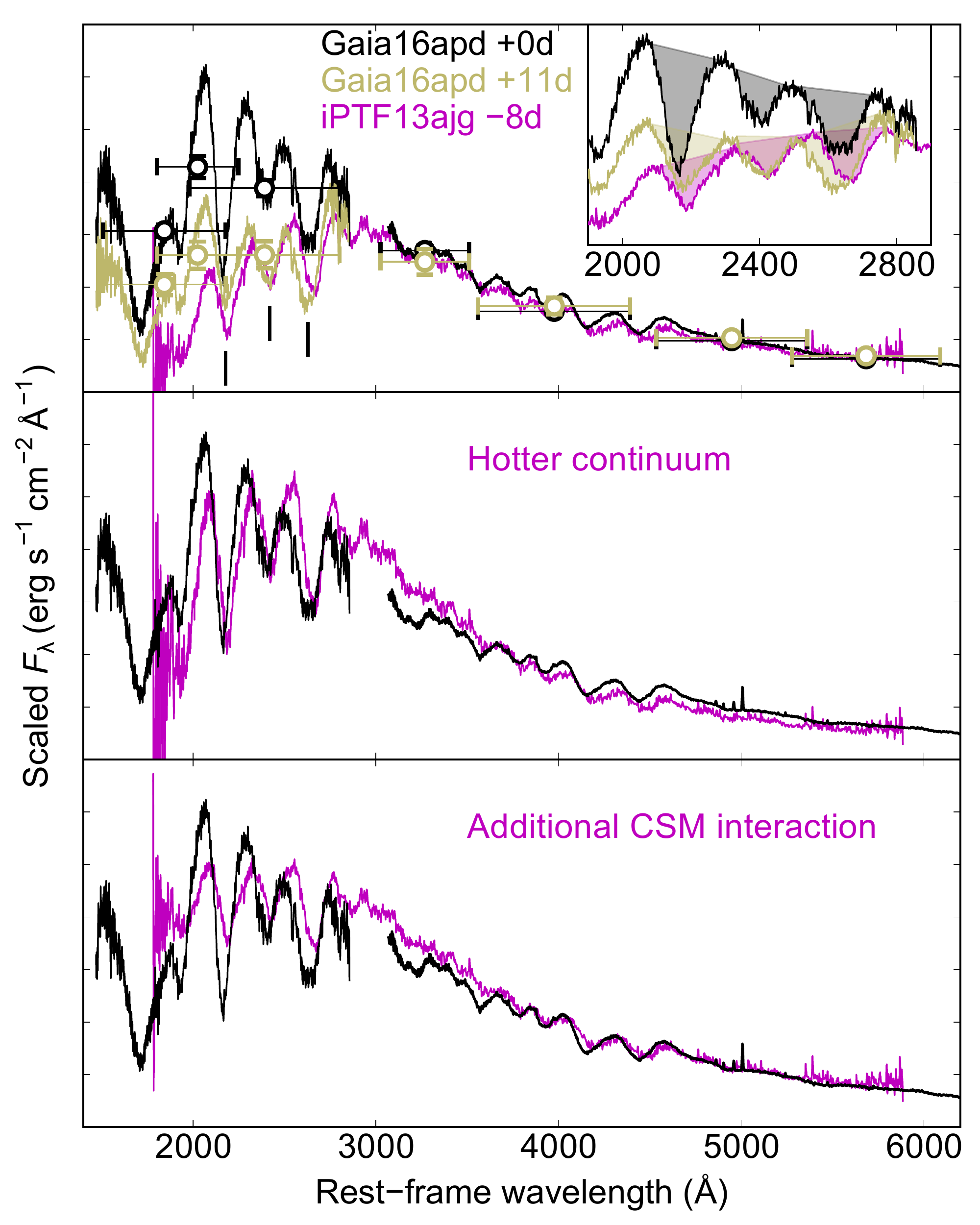}
\caption{\small {Top: Early UV-optical spectra of Gaia16apd compared with iPTF13ajg. Lines used for equivalent width comparison are marked; we measure very similar values for these three spectra, indicating that Gaia16apd is not lacking in line absorption relative to other SLSNe. Middle: rescaling the spectrum of iPTF13ajg to a hotter continuum can reproduce the spectrum of Gaia16apd. Bottom: adding a hot thermal component to iPTF13ajg, as may be expected in an interaction scenario, cannot match the spectrum of Gaia16apd as the extra continuum dilutes the equivalent widths. This is much clearer in the UV than in the optical.}}
\label{fig:uv}
\end{figure}

In a normal SN, the spectrum arises as an approximately thermal continuum, generated by electron scattering of photons in the optically-thick interior, and is then reprocessed by absorption and scattering from atomic lines in the cooler, lower-density outer ejecta. This fast-moving ejecta leads to absorption and P Cygni lines with Doppler widths characteristic of the ejecta velocity, $\sim 10^{4}$\,km\,s$^{-1}$. The specific lines depend on the composition, ionisation and excitation of this line-forming region. Increasing the input power from a central source will generate a brighter continuum, but crucially this will still be subject to {absorption from the outer, cooler ejecta}. On the other hand, if the luminosity is generated from interaction with an external CSM, the continuum is primarily generated above the region where the broad {SN absorption} lines form, {due to the high density required by CSM models of SLSNe \citep[e.g.][]{woo2007,smi2008,des2015}}. Thus the spectrum will be a sum of the underlying SN spectrum, {with its broad lines}, and a bright, {largely featureless} thermal component {\citep[e.g.][]{ami2014}}. {If a large fraction of the luminosity is coming from interaction,} this has the unavoidable effect of diluting relative to the continuum any lines {intrinsic to the SN} \citep[e.g.][]{bra2000}, {regardless of multidimensional effects that might allow one to see through the CSM \citep[e.g.][]{ben2016}. The interaction model therefore predicts that if the underlying SLSNe are similar, the equivalent widths of SN absorption lines should be lower for greater luminosity.}

Whereas the pre-maximum optical spectra of SLSNe are largely featureless apart from the relatively shallow O \textsc{ii} lines, the UV spectra show much stronger lines of ionised silicon, carbon, magnesium and titanium \citep{qui2011,vre2014,maz2016}. If Gaia16apd is powered by interaction with external material that generates a brighter or hotter additional continuum luminosity than in other SLSNe, this should show up clearly as a decrease in the equivalent widths of these lines with respect to other events.
In Figure \ref{fig:uv}, we compare the \textit{HST} spectra with the SLSN iPTF13ajg (a spectroscopically typical event with good UV coverage, shown in Figure \ref{fig:sed}). As discussed in section \ref{sec:obs}, the early UV photometry and maximum-light \textit{HST} spectrum show that the SED turns over at $\sim 2000$\AA, compared to $\sim 3000$\AA\ in iPTF13ajg. Despite this clear difference, the spectrum of Gaia16apd displays very deep, broad UV absorption lines that strikingly match those in iPTF13ajg.

To make this more quantitative, we measured the equivalent widths of the three strongest near-UV lines (marked on Figure \ref{fig:uv}) in both Gaia16apd and iPTF13ajg at the same phase relative to optical maximum. These lines have been identified by \cite{maz2016} as blends of C \textsc{ii}+C \textsc{iii}+Ti \textsc{iii} (2200\,\AA), C \textsc{ii}+Ti \textsc{iii}+Si \textsc{ii} (2400\,\AA) and Mg \textsc{ii}+C \textsc{ii} (2670\,\AA). The equivalent width is defined as $W_{\lambda} = \int(F_{\lambda}-F_{\rm cont})/F_{\rm cont} {\rm d}\lambda$, where we approximate the continuum level, $F_{\rm cont}$, as the tops of the absorption troughs (technically this is a `pseudo'-equivalent width; {\citealt{and2016}}).

For these the earliest epochs, we measure $W_{\lambda} \approx 59, 27, 51$\,\AA\ for Gaia16apd, and $W_{\lambda} \approx 58, 28, 47$\,\AA\ for iPTF13ajg. We also note that the spectrum of Gaia16apd at 11\,d after maximum (when the UV excess has largely vanished from our photometry) is a very close match to that of iPTF13ajg---and the equivalent widths of the lines in Gaia16apd do not show a significant change despite a factor $\sim 2$ change in UV luminosity (see Figure 4 inset). 
In the latest \textit{HST} spectrum of Gaia16apd at +30\,d \citep[Figure 3;][]{yan2016}, after the UV excess is gone, the equivalent widths are $W_{\lambda} \approx 59, 22, 71$\,\AA. The modest evolution is similar to that seen in iPTF13ajg over a comparable period \citep{vre2014}; we measure $W_{\lambda} \approx 59, 33, 68$\,\AA\ in the final UV spectrum available for that SLSN.
{Even if one would not necessarily expect identical line strengths due to the aforementioned caveats of composition and excitation state (the latter of which will depend on the UV flux and temperature, though the lines form out in the cooler ejecta; see also \citealt{yan2016}), the strong similarity unambiguously shows that the UV flux in Gaia16apd is not subject to significantly weaker absorption than in iPTF13ajg.}

We demonstrate this {more explicitly, and use this observation as a discriminant between physical models, in the lower panels of Figure \ref{fig:uv}. First, we re-map the spectrum of iPTF13ajg to a hotter temperature by dividing out a 15000\,K blackbody. Multiplying the normalised spectrum by a blackbody with $T \gtrsim20000$\,K gives a good match to the UV spectrum of Gaia16apd, with little effect over the optical range. By contrast, if we take the observed spectrum of iPTF13ajg and \textit{add} a hot blackbody component (again $T \sim 20000$\,K), the UV lines appear much shallower than those in Gaia16apd.} Thus the maximum-light UV spectrum of Gaia16apd can be comfortably explained by invoking the same absorption as other SLSNe relative to a hotter underlying continuum, but not by adding an external power source.

As pointed out by \citet{yan2016}, Gaia16apd emits around 50\%\ of its maximum luminosity at wavelengths below 2500\,\AA. We have just shown that this UV emission originates inside the fastest ejecta. Given that the UV luminosity alone would qualify Gaia16apd as super-luminous, this strongly favours the central engine scenario for the power source in Gaia16apd. The spectroscopic similarity to other SLSNe such as iPTF13ajg, after normalising to a hotter continuum, implies that the same mechanism is at work in the other hydrogen-poor SLSNe too. The question then is: which parameter(s) of the central engine model are able to generate a continuum in Gaia16apd that is much hotter at maximum light than almost any other event?

\section{Magnetar model and implications for SLSNe}\label{sec:mod}

\begin{figure}
\centering
\includegraphics[width=8.cm]{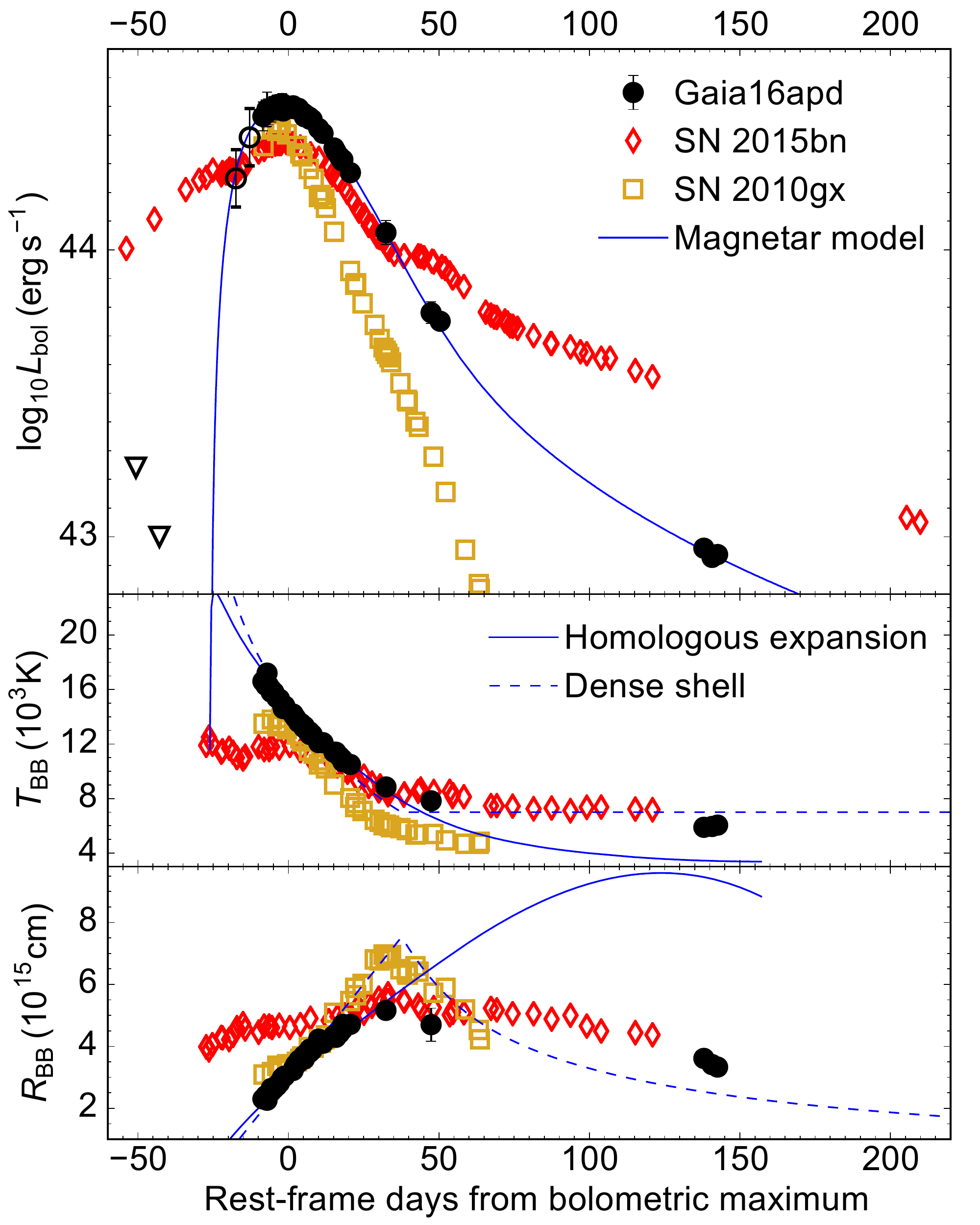}
\includegraphics[width=8.cm]{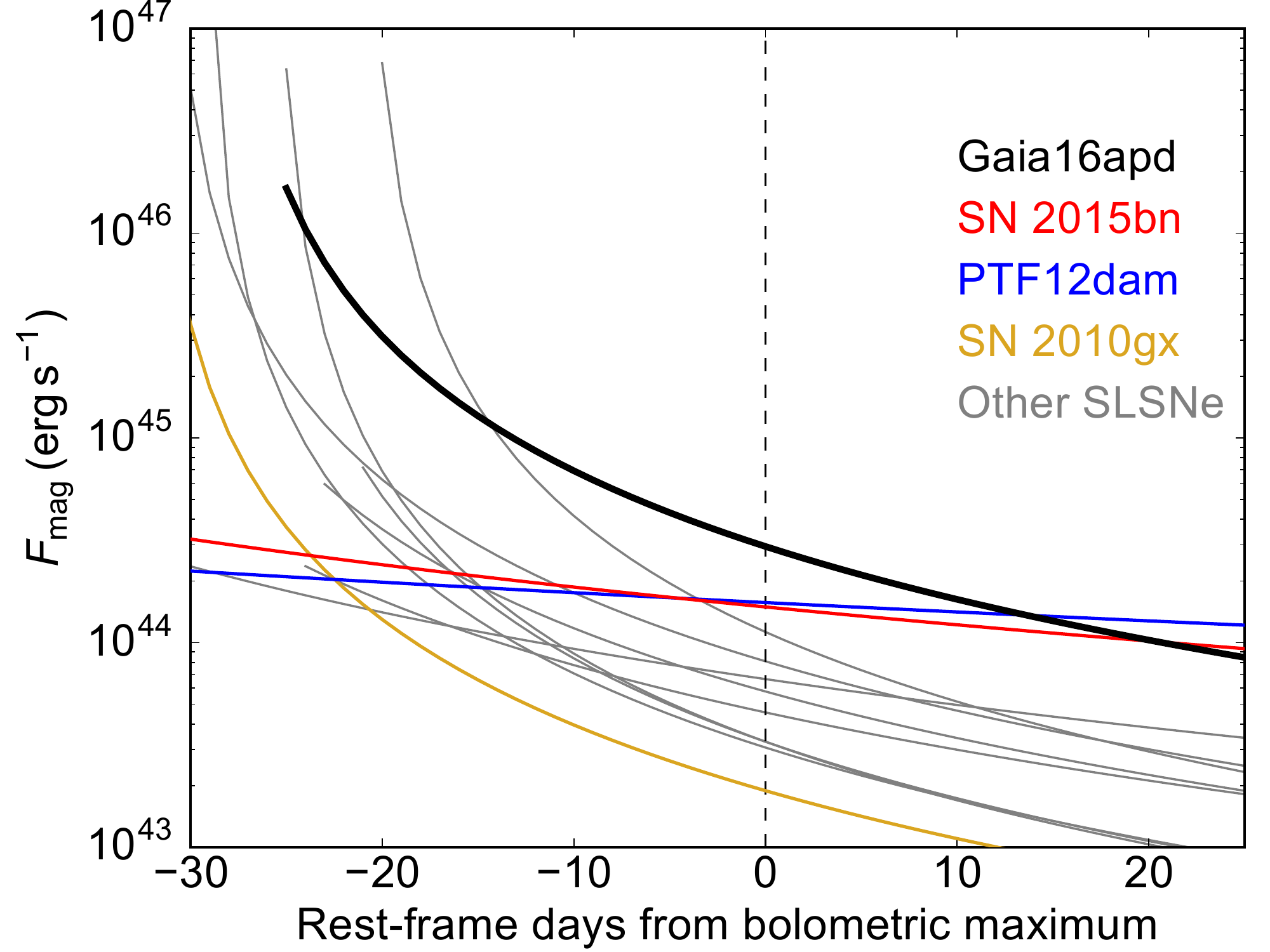}
\caption{\small Top: The bolometric light curve and blackbody temperature/radius evolution of Gaia16apd, compared with SLSNe 2015bn and 2010gx, and magnetar model fit. Bottom: The derived magnetar parameters for Gaia16apd predict a more powerful engine than any other SLSN over the period where the UV excess is observed.}
\label{fig:bol}
\end{figure}

We construct a model for Gaia16apd, assuming a millisecond magnetar as the power source. We first construct the bolometric light curve by converting our multi-band photometry to flux densities (after subtracting the host magnitudes from \citealt{yan2016}) and shifting these to the SN rest frame. We integrate over the resultant SED and approximate missing flux outside of the observed bands using blackbody fits. We assume a constant bolometric correction for the earliest two points and upper limits (these limits are from \citealt{yan2016} and \textit{Gaia}). This gives the light curve shown in Figure \ref{fig:bol}. We also plot SN\,2015bn\footnote{We correct an error in \citet{nic2016b}, where the radius in Figure 17 was too high by a factor $\sqrt\pi$} \citep{nic2016b} and SN\,2010gx \citep{pas2010}. The blackbody fits allow us to derive the evolution of the colour temperature and radius (note that this temperature is different to the suggested unabsorbed blackbodies in Figure \ref{fig:uv}), which will serve as important points of comparison for modelling.

We fit the light curve using the magnetar model first presented by \citet{ins2013}. Fixing the velocity to 10000\,\kms\ at the edge of the dense core, the photospheric velocity at maximum light is $\sim14000$\,\kms\ for their assumed density profile, in good agreement with that measured by \citet{yan2016}. We find an excellent fit to the bolometric light curve over a period of 150\,d. The best-fitting free parameters are as follows: ejected mass \Mej\,$=4.8$\,\M\ for an opacity $\kappa=0.2$\,\cmg; magnetic field $B=2.1\times10^{14}$\,G; and spin period $P=1.9$\,ms. If the opacity is instead 0.1\,\cmg, the inferred ejecta mass increases to 9.6\,\M---closer to the 12\,\M\ estimated by \citet{yan2016} for this opacity. {No leakage of $\gamma$-rays is required to fit the light curve.}

The kinetic energy in our model, $2.9\times10^{51}$\,erg\footnote{We find essentially the same energy if we fix the explosion to $10^{51}$\,erg and add a contribution from the magnetar following \citet{ins2013}.}, is smaller than that estimated by those authors, who found $E>10^{52}$\,erg. The main reason for this is that most of the ejecta behind the photosphere in our model, assuming homologous expansion, are at velocities $\ll14000$\,\kms. Given that a central engine is thought to inflate a bubble inside the ejecta \citep{kas2010,woo2010}, homology may be a coarse assumption. The true kinetic energy likely resides somewhere between our estimate and that of \citet{yan2016}, but seems to be at least a few times $10^{51}$\,erg---higher than in a canonical SN.

The model gives a good match to the radius and effective temperature of Gaia16apd until about a month after maximum light. The fit becomes poor at much later times, when the measured temperature and radius are relatively constant compared to the model. This could be attributable to the simplicity of our assumed density distribution. {We demonstrate this in Figure \ref{fig:bol} by using an alternative model for the photosphere, where it expands at constant velocity, without recession in mass coordinate, before eventually receding rapidly after cooling to a fixed temperature (taken to be 7000\,K to match the late-time observations). Although simplistic, this `dense shell' model gives an equally good match to the temperature and radius over the first 30\,d, and a better fit beyond 50\,d.}

The physical reason for the high temperature is revealed by our fit parameters: Gaia16apd falls in a seemingly unusual `sweet spot', as we will now demonstrate.
Following \citet{kas2010}, the magnetar spin-down timescale is 
\begin{equation}
\tau_{\rm sd}=4.7 (B/10^{14}\,{\rm G})^{-2} (P/{\rm ms})^2\,{\rm d,}
\end{equation}
and the power function, which is the heating term in the light curve integral \citep{arn1982}, is given by
\begin{equation}
F_{\rm mag}(t) = 4.9\times10^{46} \left(\frac{B}{10^{14}\,\rm{G}}\right)^2  \left(\frac{P}{1\,\rm{ms}}\right)^{-4} \left(1 + \frac{t}{\tau_{\rm sd}}\right)^{-2}.
\end{equation}
For a given \tsd, the early-time power input is maximised by minimising the spin period\footnote{At later times, $t\gg$\,\tsd, the dependence on $P$ cancels out, and equations 1 and 2 together yield $F_{\rm mag}\propto B^{-2}$. Diversity in $B$ thus gives the fast and slowly-declining types of SLSNe \citep{nic2013}.}. However, more powerful magnetars also spin down more quickly according to equation 1.

We compile a comparison sample of all low-redshift SLSNe that have been fitted with the same magnetar model in the literature: SN\,2015bn, PTF12dam, SN\,2010gx, PS1-11ap \citep{mcc2014}, the 5 objects from \citet{ins2013}, and the 3 objects from \citet{nic2014}. 
We observe that in all SLSNe with magnetar model fits, those with the long rise times (typically more massive ejecta) tend to have $B\approx 10^{14}$\,G, which is a factor of $\sim4$--8 weaker than in the fast-rising (lower-mass) events. 
Gaia16apd is unusual in that it has a rise time of 26\,d (at the short end for SLSNe; \citealt{nic2015b}) in combination a fairly low magnetic field, $B=2\times 10^{14}$\,G. This gives a longer \tsd\ (5.6\,d) than other SLSNe of comparable rise time. The short spin period of 2\,ms (the physical lower limit is $\approx1$\,ms; e.g.~\citealt{met2015}) is faster than in some events, though many others have a comparable period.

Using the observed rise times and the fitted $B$ and $P$, we use equation 1 to determine $F_{\rm mag}(t)$ for each SLSN (bottom of Figure \ref{fig:bol}). For around 2 weeks at either side of maximum light, Gaia16apd has a power source that is $\ga2$ times more energetic than any other event. The UV spectral models for SLSNe presented by \citet[][their Figure 11]{how2013} show how, for all other things equal, a more powerful central source gives a bluer UV spectrum exactly as we observe. \citet{nic2016b} pointed out that slowly-fading SLSNe like SN\,2015bn stay blue for much longer than other SLSNe; this is consistent with the fact that beyond $\sim20$\,d after maximum light SN\,2015bn and PTF12dam have the largest $F_{\rm mag}$.

The early UV excess in Gaia16apd also owes to the short rise time. Figure \ref{fig:bol} shows that the ejecta are still relatively compact at maximum light (e.g.~compared to SN\,2015bn); if we assume an underlying blackbody SED then a higher temperature naturally follows from injecting this much energy into a modest radius. After a few weeks from maximum light, both the measured radius and inferred engine power are similar between Gaia16apd and SN\,2015bn, consistent with the convergence in their colour evolution (Figure \ref{fig:phot}).

It is important to remember that this analysis has stemmed simply from a magnetar model fit to the bolometric light curve of Gaia16apd, with no constraints on the colours. That the derived parameters give a complete and straight-forward explanation for the UV excess (and in fact the relative colour evolution of both fast and slow SLSNe) therefore constitutes strong evidence in favour of the magnetar model. It is also possible that another type of engine could generate the same behaviour \citep[e.g.][]{dex2013,gil2016}, but we suggest that any such model would require two free parameters to set the engine's luminosity and timescale (in addition to the ejecta mass) in order to simultaneously reproduce the diversity in bolometric and colour evolution in the SLSN population.

\section{Conclusions}

Despite an apparently normal evolution in the optical, Gaia16apd is the most UV-luminous SLSN yet discovered (excepting the controversial ASASSN-15lh). While low {metal abundance} likely is a factor in the overall UV-brightness of SLSNe relative to normal SNe, as suggested by \citet{yan2016} and \citet{maz2016}, this alone does not account for the diverse UV luminosities within the SLSN class, since Gaia16apd shows the same degree of absorption as other objects that are much less luminous in the UV.
In fact, the equivalent widths of the UV absorption lines and their evolution compared to other SLSNe seem to necessitate a powerful central energy source.

Building a model for the light curve, we showed that one can self-consistently explain both the overall luminosity and the UV excess of Gaia16apd in a magnetar-powered explosion.
The key properties are a \textit{short spin period} setting a high overall energy scale, relatively \textit{low mass} giving a short rise time and thus a smaller radius (corresponding to a higher temperature) at maximum light, and most importantly a \textit{weaker magnetic field} than any other fast-rising event, increasing the spin-down time so that more power is injected around peak. 

Taking this result along with other recent observational progress, such as the link between the nebular spectra of SLSNe and gamma-ray burst SNe \citep{nic2016c,jer2016b}, and their similar geometry \citep{ins2016b}, as well as theoretical work in magnetar formation \citep{mos2015}, it now seems clear that a central engine---most likely a millisecond magnetar---is the power source in hydrogen-poor SLSNe.

Another important implication of these results is the need to follow up all SLSNe at UV wavelengths. UV data for nearby SLSNe is still fairly sparse, yet clearly the optical data for Gaia16apd told only part of the story. The discovery that some SLSNe are this bright in the UV is a major boost for future high-redshift searches \citep[see also][]{ins2014,yan2016}. JWST should easily detect Gaia16apd-like events at $z\ga10$ as approximately year-long near-infrared transients (unlike pair-instability SNe, which would have much longer timescales and should be faint in the rest-frame UV), offering perhaps the most promising opportunity yet to observe the deaths of the first stars {with upcoming optical and NIR instruments }.

\acknowledgments
R.M. acknowledges generous support from NASA Grant NNX16AT81G.

\bibliographystyle{apj}

\bibliography{/Users/matt/Documents/Papers/mybib}

\end{document}